\documentclass[usenatbib,amsmath,amssymb,amsbsy,onecolumn]{mn2e}

\usepackage{natbib}
\usepackage{graphicx}
\def\X{{\mathrm{x}}}
\def\Y{{\mathrm{y}}}

\def\n{{\rm n}}
\def\p{{\rm p}}
\def\e{{\rm e}}
\def\s{{\rm s}}

\def\L{{\rm L}}

\def\be{\begin{equation}}
\def\ee{\end{equation}}
\def\bea{\begin{eqnarray}}
\def\eea{\end{eqnarray}}

\newcommand{\pd}[2]{\frac{\partial {#1}}{\partial {#2}}}

\begin{document}

\title{Superfluid neutron star turbulence}

\author[Andersson, Sidery \& Comer]{N.~Andersson$^1$, T. Sidery$^1$  and G.L. Comer$^2$ \\
$^1$ School of Mathematics, University of Southampton, 
Southampton SO17 1BJ, United Kingdom\\
$^2$ Department of Physics \& Center for Fluids at All Scales, 
Saint Louis University, St.~Louis,
MO, 63156-0907, USA}

\maketitle

\begin{abstract}
We analyse the implications of superfluid turbulence for neutron star physics. 
We begin by extending our previous results for the mutual friction force for a straight vortex array to account for the self-induced flow which arises when the 
vortices are curved. We then discuss Vinen's phenomenological model for isotropic turbulence, and derive the associated (Gorter-Mellink)
form for the mutual friction. We compare this derivation to a more recent analysis of
Schwarz, which sheds light on various involved issues. Having discussed isotropic turbulence, we argue that 
this case is unlikely to be relevant for neutron stars. Instead we expect a rotating neutron star to exhibit polarised turbulence, 
where relative flow drives the turbulence and rotation counteracts it. Based on recent results for superfluid Helium, we 
construct a phenomenological model that should have the key features of such a polarised turbulent system.  
\end{abstract}

\section{Introduction}

Turbulence in superfluid Helium has been exciting 
experimenters and theorists in low temperature physics
for the last 50 years. Not surprisingly, 
the literature on the subject is vast, and
the research has  reached the level where 
theoretical models represent  experimental data 
relatively well \citep{donnelly}. Meanwhile, the possibility that 
turbulence may be relevant for neutron star superfluidity 
has hardly been discussed at all (although see the recent work by \citet{peralta1,peralta2}). 
This is quite surprising
because the standard models for neutron stars are 
strongly inspired by the two-fluid model for superfluid Helium. 
There may be  differences between the two problems, but the
underlying multifluid dynamics is essentially the same \citep{mendell1,prix,monster}.
One would expect this to be particularly true for the vortex dynamics
and the processes that lead to vortex tangles and 
turbulent behaviour.

The aim of this paper is to discuss turbulence in 
superfluid neutron stars. In particular, we want to extend the description 
of the mutual friction force  \citep{als,mendell2,trevor}, that couples the two components in a 
superfluid neutron star core, in such a way that we can account for 
vortex tangles. This is an important first step along a road that 
should eventually lead to an improved description of 
superfluid neutron star dynamics. The mutual friction coupling 
enters in the analysis of three important astrophysics problems. 
First of all, it sets the timescale on which the two core fluids are 
coupled following a pulsar glitch (or any event that involves a velocity 
difference) \citep{as88,jahanmiri,trevor}. Secondly, it is one of the key mechanisms that 
determine the damping rate of neutron star free precession \citep{jones}. 
Finally, the mutual friction is one of the main damping agents
for superfluid neutron star oscillations. In particular,    
it has been argued that the mutual friction damping completely 
suppresses the gravitational-wave driven instability of the 
fundamental f-modes \citep{lm1}. It is also one of the key damping mechanisms affecting 
the analogous instability of the inertial r-modes \citep{lm2}, see \citet{NArev} for a review. However, all 
existing results have been obtained by making use of the 
form of the mutual friction first suggested by \citet{hallvinen}
for Helium. Since this model is based on the assumption on a straight vortex
array it is not relevant  
when the vortices form a disorganised tangle. The mutual friction 
force that applies in this case is known to be rather different \citep{vinen57}.
This suggests that, if neutron stars exhibit turbulent behaviour,
then much of the generally accepted thinking about the interior superfluid
dynamics may be wrong.  

\section{The straight vortex case}

We have recently discussed the mutual friction force, and 
the associated coefficients, that should apply when the 
vortices are straight \citep{trevor}. This should be the relevant case for 
uniformly rotating systems, provided that the vortex array is dynamically 
stable (which may not be the case).  Our analysis was based on 
the standard two-fluid model \citep{prix,monster} where the superfluid neutrons, 
with mass density $\rho_\n$ and velocity $v^i_\n$, are distinguished from a 
conglomerate of all charged components, represented by 
$\rho_\p$ and $v^i_\p$ and loosely refered to as the ``protons''.
In absence of dissipation the equations of motion for the 
two fluids can be written
\begin{eqnarray}\label{euler}
\left(\pd{}{t} + {v}^{j}_{\X}\nabla_{j} \right)   
\left[ {v}_{i}^{\X} + \varepsilon_{\X} w_i^{\Y\X} \right] + \nabla_{i} (\Phi + 
\tilde{\mu}_{\X}) 
+ \varepsilon_{\X}  w_j^{\Y\X} \nabla_{i} v^{j}_{\X} = 0 \ .
\end{eqnarray}
where the contituent indices $\X$ and $\Y\neq\X$ can be either $\n$ or $\p$.
The relative velocity is defined as
\be
w_i^{\Y\X} =  {v}_{i}^{\Y} - {v}_{i}^{\X} \ . 
\ee
Furthermore,
\be 
\tilde{\mu}_\X = {\mu_\X \over m_\X} = { 1 \over m_\X} { \partial E \over \partial n_\X} \ ,
\ee
where $E$ is an energy functional (which represents the equation of state in this formalism) 
and $\tilde{\mu}_\X$ is the relevant chemical potential per unit 
mass.  The  coefficients
\be
\varepsilon_\X = {2 \alpha \over \rho_\X}  \qquad  \mbox{ where } \qquad \alpha = { \partial E \over \partial w_{\n\p}^2} \ ,
\ee 
encode what is known as the entrainment effect, 
and
$\Phi$ represents
the gravitational potential.
For a detailed discussion of these equations, see \citet{prix} and \citet{monster}.
Since the equations of motion represent momentum balance, we
see that the momentum per fluid element is
\be
p_i^\X =  m_\X[ {v}_{i}^{\X} + \varepsilon_{\X} w_i^{\Y\X}]  \ , 
\label{mom_def}\ee
This illustrates the main impact of the entrainment. The momentum of the 
each fluid need not be parallel with that fluids transport velocity.
This effect is important for neutron star cores because the strong force
endowes each neutron with a cloud of protons (and vice versa). 
This affects the effective mass of the particle and, when it moves, alters 
the momentum. As discussed by, for example, \citet{trevor} the entrainment can be related to the 
effective proton mass $m_\p^*$ via
 \be
2 \alpha = \rho_\p \varepsilon_\p =  n_\p (m_\p - m_\p^*) = n_\p \delta m_\p^* \ .
\ee

Formally, the above equations represent a
macroscopic average over a large number of individual fluid elements (in the standard way).
This is a natural approach for neutron stars where one is interested 
in fluid dynamics at lengthscales vastly larger than the inter-particle separation. 
However, in order to assign values to the various parameters (like the entrainment $\alpha$)
one must regularly resolve the problem at the microscopic level. This then raises
issues concerning the averaging procedure which is required to make contact with 
the macroscopic level. In problems involving vortex dynamics, the situation is made 
even more complex. It is then natural to introduce an intermediate ``mesoscopic'' 
level which is sufficiently large that one can average over a large collection of 
particles (and hence discuss ``fluid'' dynamics in a sensible way) and yet small
enough that one can resolve the individual vortices. The macroscopic equations then 
derive from an averaging over a large set of vortices, leading to the smooth 
equations of motion  (\ref{euler}). 
As we will see, this averaging is one of the key problems that must be overcome 
in a description of superfluid turbulence. 

On the mesoscopic level, the motion of a neutron vortex is affected by 
fluid flow past it (via the Magnus ``force''). At the same time, 
the circulation associated with the vortex induces circular flow in the 
protons because of the entrainment. This in turn generates a magnetic field on
the vortex, and the scattering of electrons off of this field 
acts resistively on the vortex motion \citep{als}.
Taking this  resistive force to be proportional to the difference in velocity between the vortex ($v_\L^i$) and  the charged fluid flow
we have
\begin{equation}
f_{i}^\mathrm{e} = {\cal R} (v^{\p}_{i} - v^{\L}_{i}) \ .
\label{resist}\end{equation}
acting per unit length of the vortex.
Neglecting the inertia of the vortex core,
and balancing $f_i^\mathrm{e}$ to the Magnus ``force'' we arrive at the mutual friction  
acting on the neutrons (per unit length of vortex)
\be
    f^{\rm mf}_i = {\cal B}^\prime \rho_\n \epsilon_{i j k} \kappa^j 
                   w_{\n\p}^k + {\cal B} \rho_\n \epsilon_{i j k} 
                   \epsilon^{k l m} \hat{\kappa}^j \kappa_l w^{\n\p}_m 
                   \ , \label{fmf1} 
\ee
(correcting an overall sign error resulting from  \citet{trevor}
writing the consitutent indices in the wrong order). The same force acts on the vortex and an equal and opposite force 
affects the protons. The vector $\kappa^i$ determines the orientation of 
the vortex (see later) and its magnitude (the quantum of circulation) 
is $\kappa = h/2m_\n\approx 2\times 10^{-3}$~cm$^2$/s.
Our estimated (dimensionless) mutual friction coefficients 
are 
\be
{\cal B} = { { \cal R} \over \rho_\n \kappa} \approx  
4 \times10^{-4} \left( {\delta m_\p^* \over m_\p} \right)^2
\left( {m_\p \over m_\p^*} \right)^{1/2} \left( { x_\p \over 0.05} \right)^{7/6}  \left( { \rho \over 10^{14} \mathrm{g/cm}^3} \right)^{1/6} \ ,
\ee
and
\be
{\cal B}^\prime = {\cal B}^2\ , 
\ee
where $x_\p = \rho_\p/\rho$ is the proton fraction, 
in agreement with previous estimates by \citet{als} and \citet{mendell2}. 

In order to complete the description of the mutual friction force we need to note that  
the  equations of motion (\ref{euler}) are relevant on the macroscopic scale. 
Meanwhile the analysis leading to (\ref{fmf1}) is mesoscopic, as we are 
considering a single vortex. The macroscopic equations of motion follow if
we average over a large collection of vortices, i.e. we need to work at a sufficiently
large lengthscale that such averaging is meaningful. We also need to 
note that the Magnus effect is already accounted for in the left-hand side
of  (\ref{euler}), which means that we only need to add the smooth averaged
form of the force (\ref{fmf1}) to the right-hand side.
In the particular case of a straight vortex array, the required 
averaging is easy. To get the required force per unit volume we 
simply need to multiply (\ref{fmf1}) by the local number density of vortices 
per unit area, $n_v$. 

As we will face this issue again later, it is useful to
discuss it in a little bit more detail here.    
At the fluid dynamics level,  we introduce the 
``rotation velocity'' in the standard way, 
\be
2 \Omega_\X^i = \epsilon^{ijk} \nabla_j v_k^\X \ ,
\ee 
This means that, on the macroscopic level we identify (assuming that the lengthscale considered is
sufficiently small that we can treat $\varepsilon_\n$ as constant)
\be
\epsilon^{ijk} \nabla_j p^\n_k = 2 m_\n [ \Omega_\n^i + \varepsilon_\n (\Omega_\p^i - \Omega_\n^i)] \ .
\label{mac}\ee
Meanwhile, at the mesoscopic level the circulation of neutron momentum (not velocity) is quantised. 
Representing the mesoscopic ``momentum''\footnote{We have chosen to refer to the mesoscopic quantity $\bar{p}_\n^k$
as the ``momentum'', even though this is not dimensionally correct. As should be clear from the discussion 
we obtain the true momentum by averaging over a unit area. A similar comment applies to the discussion following 
eq.~(\ref{mesos}).} 
 by $\bar{p}_\n^k$, we have 
\begin{equation}
{1 \over m_\n} \epsilon_{ijk}\nabla^{j} \bar{p}_\n^k = \kappa_i = { h \over 2 m_\n} \hat{\kappa}_i \ .
\label{mes}\end{equation}
Strictly speaking, we should arrive at the first equation by averaging the second. Denoting the average
by angular brackets we  have (in the case of straight vortices) 
\be
\langle \epsilon^{ijk} \nabla_j \bar{p}^\n_k \rangle = n_v m_\n \kappa^i = \epsilon^{ijk} \nabla_j p^\n_k
\label{average1}\ee
Hence we see that 
\be
n_v \kappa = 2  [ \Omega_\n + \varepsilon_\n (\Omega_\p - \Omega_\n)] \ .
\label{n_v}\ee

\section{Are superfluid neutron stars likely to be turbulent?}

We want to extend our description of the mutual friction force to the more complicated situation where the 
vortices are curved and may reconnect to form a turbulent tangle. Before we 
try to do this, it is useful to have some guidance as to whether 
one would expect turbulence to be relevant for superfluid neutron stars. 
After all, virtually every discussion of neutron star vortex dynamics has made
the assumption that the vortex array is straight. 
Should this basic tenet fall, a possibility that has been hinted at by, for example, 
\citet{packard} and \citet{chevalier}, we might have to reassess much of our current 
``wisdom''. 

To get an idea of the possible relevance of superfluid turbulence, 
let us adapt a recent argument due to \citet{finne}. They consider the problem for 
superfluid $^3$He, and argue that an intrinsic parameter
determines the onset of turbulence. In our case, the argument would 
proceed as follows. First assume that the motion of the ``normal fluid'' can be neglected,
i.e. take $v_i^\p=0$.  
This is convenient becase we then do not have to worry about the 
effects of entrainment. In the Helium case it may also be quite reasonable 
because one can assume that the normal fluid is viscous and essentially immobile with respect to the 
walls of the container (at least on a timescale large compared to the viscous timescale),
In our case, the equation of motion for the neutrons can then be written
\be
(\partial_t + v_\n^j \nabla_j ) v_i^\n + \nabla_i [ \ldots ] = { 1 \over \rho_\n} f^\mathrm{mf}_i
\ee
This is easily rewritten as
\be
\partial_t v_i^\n  + \nabla_i [ \ldots ] = \epsilon_{ijk}v_\n^j \omega_\n^k +  { 1 \over \rho_\n} f^\mathrm{mf}_i
\ee
where $\omega_\n^i= \epsilon^{ijk}\nabla_j v^\n_k$, Taking the curl of this equation, and using (\ref{fmf1}),
we arrive at (assuming that the various parameters are constant)
\be
\partial_t \vec{\omega}^\n = ( 1 - \mathcal{B}^\prime) \nabla\times( \vec{v}_\n\times \vec{\omega}_\n) 
+ \mathcal{B} \nabla \times [ \hat{\omega}_\n \times ( \vec{\omega}_\n \times \vec{v}_\n)]
\ee
Comparing to the standard fluid case, \citet{finne} argues that the 
 first term on the right hand side (the inertial term) induces the transfer
of energy to smaller length scales, i.e. drives the turbulence. 
Meanwhile, the second term leads to damping that serves to stabilise the flow.
The relative importance of the two effects is determined 
by the parameter
\be
q = { \mathcal{B} \over 1 - \mathcal{B}^\prime}  
\ee
If $q<1$ one might expect the flow to become turbulent. This is
in quite good agreement with the experimental evidence provided by \citet{finne}.
They show that turbulence sets in for $q\le 1.3$ or so. 
Further quantitative evidence for this idea has been provided by \citet{henderson}.
They analyse the spectrum of Kelvin waves in the vortex array, and argue that 
$q=1$ represents the crossover point at which these waves change from being overdamped
(damping out before completing a full oscillation) to propagating.  The idea
would be that the vortex oscillations can then lead to reconnections with neighbouring vortices. 
This is obviously more likely to happen if the waves propagate along the   
vortex array. Anyway, in the neutron star case we have already shown that 
$\mathcal{B}^\prime \approx \mathcal{B}^2 \ll 1$. Hence, we would always have
$q\ll 1$, and according to the above criterion one would expect a 
superfluid neutron star core to be extremely 
susceptible to becoming turbulent. 
Of course, there are a number of caveats. Most importantly, 
the argument does not  account for the stabilising effect of 
rotation. Nevertheless, it is clear that we must take the possibility 
of superfluid turbulence seriously.   

\section{Accounting for vortex curvature} 

The mutual friction force (\ref{fmf1}) is only relevant when the
vortices are straight. This should be the case in a star 
that is uniformly rotating, assuming that the vortex array is dynamically 
stable. This has generally been taken to be the case for neutron stars.
However, work on the analogous problem for superfluid Helium indicates that 
there will exist a critical relative 
flow above which oscillations are induced in the vortices rendering
the array 
unstable \citep{glaberson}. 
This leads to the formation of a complex vortex tangle and a state 
of turbulence. As a first step towards understanding the implications of 
this for neutron stars we will generalise the mutual friction force
to the case of isotropic turbulence. 
To do this, we need to move away from the assumption that the vortices 
are straight.  

The induced flow due to 
the vortex curvature will play a key r\^ole in our discussion. 
The analysis of this problem for 
superfluid Helium \citep{donnelly} is easily extended to the neutron star case. This
is not surprising since
the basic nature of the rotational vortices is the same. Nevertheless, 
the derivation is provided in Appendix~A. It should be useful for readers that 
are unfamiliar with the Helium results. Moreover, our analysis accounts for
the entrainment effect which has (as far as we know) not 
been considered in previous work. Yet we know that it is important
for neutron star dynamics. In fact, one would not be at all surprised if 
(at the end of the day) it were relevant for the Helium problem as well.

As demonstrated in Appendix~A, the vortex curvature induces a contribution to the 
(mesoscopic) neutron momentum
\be
{\bar{p}^\mathrm{ind}_i \over m_\n} = \nu \epsilon_{ijk}s^{\prime j} s^{\prime\prime k} 
\label{mesos}\ee
where
\be
\nu = { \kappa \over 4 \pi} \log \left( { b \over a_0} \right) 
\ee
and we also know that 
\bea
s^\prime_i &=& \hat{\kappa}_i \\
s^{\prime\prime}_i &=& \hat{\kappa}^j \nabla_j \hat{\kappa}_i 
\eea
To get an idea of the value of the parameter $\nu$, which determines the tension in the vortex array, 
we note that a typical value of the size of the vortex core would be $a_0 \approx 100$~fm, while 
we can let $b$ be given by the inter-vortex spacing. This would mean that (for the usual triangular
vortex configuration) we would have
\be
b \approx 3.4 \times 10^{-3} \left( { \Omega_\n \over 100 \mbox{ rad/s} } \right)^{-1/2} \mbox{ cm}
\ee
Combining these we see that
\be
\log \left( { b \over a_0} \right) \approx 20 - { 1 \over 2} \log \left( { \Omega_\n \over 100 \mbox{ rad/s} } \right)
\ee
In other words, for superfluid rotation rates at the opposite ends of the range of observed pulsar periods, $P=10^{-3}-10$~s (say) 
the value of $\nu$ only varies by about 30\%. It is also worth noting that $\nu$ is similar in the Helium case. 
Typical values would then be $a_0\approx10^{-8}$~cm and $b\approx 10^{-2}$~cm, leading to $\log(b/a_0) \approx 14$.

Having derived the curvature-induced momentum for a single vortex, we can apply the result to the two-fluid model 
for a neutron star core. As discussed  by \citet{trevor}, the presence of a neutron vortex 
will affect the flow of both  neutrons and  protons. The induced flow makes a contribution
\bea
{\bar{p}_i^\n \over m_\n} &=& v_i^\n + \varepsilon_\n ( v_i^\p - v_i^\n)  =  {\bar{p}^\mathrm{ind}_i \over m_\n} \\
{\bar{p}_i^\p \over m_\p} &=& v_i^\p + \varepsilon_\p ( v_i^\n - v_i^\p)  = 0
\eea
Solving these equations for the individual velocities, we find
\begin{equation}
v^{\n}_{i} = \frac{1 - \varepsilon_{\p}}{1- \varepsilon_{\n} - \varepsilon_{\p}}  \nu \epsilon_{ijk}s^{\prime j} s^{\prime\prime k} 
\end{equation}
\begin{equation}
v^{\p}_{i} = - \frac{ \varepsilon_{\p}}{1- \varepsilon_{\n} - \varepsilon_{\p}}\nu \epsilon_{ijk}s^{\prime j} s^{\prime\prime k} 
\end{equation}
That is,  the  difference between the two induced velocities is
\begin{equation}
v^{\n}_{i} - v^{\p}_{i} = \tilde{\nu}  \epsilon_{ijk} s'^{j} s''^{k}
\label{veldiff}
\end{equation}
where we have defined 
\be
\tilde{\nu} = \frac{1}{1- \varepsilon_{\n} - \varepsilon_{\p}} \nu
\ee

How do these additional contributions to the flow affect the mutual friction force? The answer is quite intuitive. 
The induced velocities simply add to the general flows $v_\X^i$ in the expression for the force (\ref{fmf1}).
Adding the velocity difference (\ref{veldiff}) to the overall relative velocity in (\ref{fmf1}),  and noting that
\be
\vec{s}^\prime \times (\vec{s}^\prime \times \vec{s}^{\prime\prime}) = - \vec{s}^{\prime\prime} \ , 
\ee 
we find that the additional contribution to the mutual friction force can be written
\bea
f^{\rm ind}_i =  - \rho_\n \kappa \tilde{\nu}  \left[ {\cal B}^\prime  s_i^{\prime\prime} + 
                  {\cal B} \epsilon_{ijk} s^{\prime j} s^{\prime \prime k} \right]
                   \ , 
\label{find}\eea
or equivalently
\be
f^{\rm ind}_i =  - \rho_\n \kappa \tilde{\nu}  \left[ {\cal B}^\prime \hat{\kappa}^j \nabla_j \hat{\kappa}_i 
+    {\cal B} \epsilon_{ijk} \hat{\kappa}^j \hat{\kappa}^l \nabla_l \hat{\kappa}^k \right]
\ee
To get an expression for the total mutual friction force, these terms should be added to (\ref{fmf1}).
Thus we get the force per unit length
\be
    f^{\rm mf}_i = {\cal B}^\prime \rho_\n \epsilon_{i j k} \kappa^j 
                   w_{\n\p}^k + {\cal B} \rho_\n \epsilon_{i j k} 
                   \epsilon^{k l m} \hat{\kappa}^j \kappa_l w^{\n\p}_m
- \rho_\n \tilde{\nu}  \left[ {\cal B}^\prime \hat{\kappa}^j \nabla_j \kappa_i 
+    {\cal B} \epsilon_{ijk} \kappa^j \hat{\kappa}^l \nabla_l \hat{\kappa}^k \right]
                   \ . \label{fmf2} 
\ee
Finally, we determine the macroscopic mutual friction by averaging over a collection of vortices. 
If we assume that the vortex curvature is sufficiently small that there are no vortex intersections, i.e. we
ignore possible reconnections and turbulent tangles, then the total force is simply arrived at by multiplying the above expression 
by $n_v$ (as in Section~2). Thus we arrive at the final result
\be
    f^{\rm mf}_i = {\cal B}^\prime \rho_\n n_v \epsilon_{i j k} \kappa^j 
                   w_{\n\p}^k + {\cal B} \rho_\n n_v \epsilon_{i j k} 
                   \epsilon^{k l m} \hat{\kappa}^j \kappa_l w^{\n\p}_m
- \rho_\n \tilde{\nu} n_v  \left[ {\cal B}^\prime \hat{\kappa}^j \nabla_j \kappa_i 
+    {\cal B} \epsilon_{ijk} \kappa^j \hat{\kappa}^l \nabla_l \hat{\kappa}^k \right]
                   \ . \label{mf_macro} 
\ee

\section{Turbulence}

The curvature contributions to the mutual friction  are of key importance for 
the development of superfluid turbulence. In principle, one might expect the above result to remain valid
(to some extent) also when the vortices are in a tangle. At least as long as the separation between different 
vortex segments is large enough that the local analysis leading to (say) the induced flow remains valid, and as long as it is legitimate
to ignore reconnections.  However, when the vortices are in a tangle 
it is far from easy to extend the mesoscopic result to the macroscopic fluid dynamics level.
The key issue concerns the averaging procedure. In the case of a straight 
vortex array, this issue was trivial. We simply needed to know the number of vortices 
per unit area. Once we allow for a tangle it is no longer obvious how 
one should average over a collection of vortices. This difficulty was recognized already by Feynman
in his seminal description of superfluid turbulence \citep{feynman}. One has to be wary of the need to account for 
vortex loops on a scale smaller than that of the ``fluid elements''. These vortex loops  need to be 
accounted for even if the ``resolution'' of the averaged equations is too coarse to identify them. 
There are a number of seriously challenging questions here. 

To make some progress we adopt a phenomenological approach, 
following the pioneering work of \citet{vinen57}. We will then contrast the results with a more recent analysis
due to \citet{schw82,schw88}. The results of the two approaches agree qualitatively, even 
if there are differences at the level of interpretation.

\subsection{Vinen's approach}

In one of his many key papers in this area, \citet{vinen57} discussed superfluid turbulence and the 
associated mutual friction force. He argued that, in the case of an isotropic
vortex tangle the mutual friction should be proportional to the cube of the relative velocity in the flow.  
This general form for the force had previously been postulated by \citet{gorter}.
In this section we will reproduce Vinen's  
argument for the neutron star problem. 
This leads to an expression for the mutual friction force that can be used when the 
neutron superfluid is in a homogeneous turbulent state. The analysis is essentially phenomenological, but it
has the advantage that it leads to an explicit expression for the force in terms of the parameters from 
the straight vortex case, cf. Eq.~(\ref{fmf1}).

We start from the assumption that the resistive force (\ref{resist}) still applies  in the case of a vortex tangle. This 
seems reasonable
provided that the vortex segments remain well separated, and we focus 
on a local part of the vortex\footnote{Caveats: It is not clear to what extent this key assumption is justified. One must worry about the r\^ole of reconnections in the vortex 
tangle. It is also not obvious that the resistivity parameters remain as in the straight vortex case. In the neutron star problem one 
would need to consider the local magnetic field induced by entrained protons flowing around the vortex, see \cite{trevor}, and the 
resultant scattering of electrons flowing through a vortex tangle rather than a straight vortex array. Nevertheless, it is natural to make this assumption
in this first analysis 
of the problem.} . Let $L$ be the total vortex line length per unit volume
and assume that the vortex tangle moves (on average) with the superfluid, i.e. use $\langle v_i^\L \rangle \approx v_i^\n$. 
In the case of isotropic turbulence we then have a force per unit volume
\citep{vinen57}
\be
f_i^\mathrm{mf} =  { 2 L \over 3} {\cal R} ( v_i^\p - v_i^\n)
\label{fturb1}\ee

To estimate $L$, Vinen balances the growth rate due to the Magnus effect (which in turn balances $f_i^\e$ in the 
equation of motion for the vortex) to dissipation. To get the growth rate one assumes that it is proportional to 
(\ref{resist}), i.e.  use $f_e\approx {\cal R} w_{\p\n} = \rho_\n \kappa {\cal B} w_{\p\n}$. 
Then the fact that it should in addition only depend on $\rho_\n$, $L$ and $\kappa$,
together with a dimensional analysis\footnote{Note: One could allow additional dimensionless combinations of the various parameters to 
enter here. Also, because the argument is based on a dimensional analysis, it is not easy to see how the dimensionless 
entrainment would enter.} leads to
\be
{dL_+ \over dt} = { \chi_1 f_\e L^{3/2} \over \rho_\n \kappa}  
\ee
where $\chi_1$ is a dimensionless parameter (assumed to be of order unity). 
Vinen next assumes that the vortex tangle decays in the same fashion as ordinary turbulence (i.e. through a Kolmogorov cascade). 
This leads to an overall decay according to
\be
{dL_- \over dt}  = { \chi_2 \kappa L^2 \over 2 \pi} 
\ee
where $\chi_2$ is another dimensionless parameter. 
Combining the growth and the decay, we have
\be
{ dL \over dt} = {dL_+ \over dt} - {dL_- \over dt} = { \chi_1 f_\e L^{3/2} \over \rho_\n \kappa}  -  { \chi_2 \kappa L^2 \over 2 \pi} 
\ee
For obvious reasons, this has become known as ``Vinen's equation''.
We see that 
a steady state solution requires
\be
L  = \left( { 2 \pi  \over \kappa} \right)^2  \left( { \chi_1 \over \chi_2} \right)^2 {\cal B}^2 w_{\p\n}^2   
\ee 
The final expression for the turbulent mutual friction force is obtained by using this value for $L$ in 
(\ref{fturb1}). Hence, we arrive at
\be
f_i^\mathrm{mf} =  { 8 \pi^2 \rho_\n  \over 3 \kappa}   \left( { \chi_1 \over \chi_2} \right)^2  {\cal B}^3 w_{\p\n}^2 w^{\p\n}_i   
\label{vinen}\ee

This analysis leads to a mutual friction force that depends on the cube of the relative velocity $w^{\p\n}_i$. This 
behaviour was first suggested by \citet{gorter}, which is why (\ref{vinen}) is sometimes refered to as the ``Gorter-Mellink'' force. 
It is not clear how useful one should expect this representation of the turbulent mutual friction to be, 
given the various assumptions that went into the analysis. Nevertheless, it has been compared to experimental results in a variety of contexts
and appears to describe the main phenomena rather well (with typical parameters $\chi_1\approx 0.3$ and $\chi_2 \approx 1$, see Fig.~2 in \citet{vinen57}). 
This indicates that the key assumptions must be ``at least approximately true'' \citep{vinen57}.

\subsection{Schwarz's approach}

In a more recent series of papers, \citet{schw82,schw88}  used numerical simulations to investigate the vortex dynamics.
It is useful to repeat some of Schwarz's theoretical analysis here, since it highlights the issue of averaging over a vortex 
tangle. It also provides a complement to Vinen's more phenomenological discussion.

We start by recalling that 
the complete mutual friction force (per unit length of a vortex) can be written
\be
\vec{f} = \kappa \rho_\n  {\cal B}\left[  \vec{s}^\prime \times \vec{s}^\prime \times  \vec{w}_{\n\p} - \tilde{\nu} \vec{s}^\prime \times \vec{s}^{\prime\prime}
\right] +  \kappa \rho_\n   {\cal B}^\prime \left[ \vec{s}^\prime \times\vec{w}_{\n\p} - \tilde{\nu} \vec{s}^{\prime\prime}\right] 
\ee
We want to consider an isotropic turbulent state. To do this we focus on a small volume of fluid which contains a vortex tangle and through
which there is a uniform flow represented by  $\vec{w}_{\n\p}$. We will take isotropic to mean that both $\vec{s}^\prime$ and  $\vec{s}^{\prime\prime}$
are rotationally symmetric with respect to $\vec{w}_{\n\p}$. We are interested in the 
  mutual friction force per unit volume, $V$. In principle, it follows from the integral
\be
\vec{f}^\mathrm{mf} = { 1 \over V} \int \vec{f} d\xi 
\ee
where we  note that the integrand only has support on the vortex lines.
Given the symmetries of the assumed state it is easy to see that the terms proportional to ${\cal B}^\prime$ will not
give a net contribution to the integral \citep{schw88}. We also find that
\be
\int \vec{s}^\prime \times \vec{s}^\prime \times  \vec{w}_{\n\p} d\xi = -\vec{w}_{\n\p} V L I_\parallel
\ee
where we have defined
\be
I_\parallel = { 1 \over VL} \int (1 - s_\parallel^2) d\xi  
\ee
Here we have used  $\vec{w}_{\n\p} = w_{\n\p} \hat{r}_\parallel$ and $s_\parallel = \vec{s}^\prime \cdot \hat{r}_\parallel$.
To represent the induced flow term we will need
\be
I_l \hat{r}_\parallel = { 1 \over V L^{3/2} } \int  \vec{s}^\prime \times \vec{s}^{\prime\prime} d\xi
\ee
Combining the two contributions to the force we get
\be
\vec{f}^\mathrm{mf} = \kappa \rho_\n {\cal B} \left[ L I_\parallel w_{\p\n} - \tilde{\nu} I_l L^{3/2} \right] \hat{r}_\parallel
\ee
In these expressions we have used the total line length per unit volume 
\be
L = { 1 \over V} \int d\xi
\ee 
Schwarz argues that if $L$ is independent of position and system size then it should scale as
\be
L = c_\L^2 \left( {w_{\n\p} \over \tilde{\nu}} \right)^2
\ee
He further determines the scaling parameter $c_\L$ by comparing to the numerical simulations.

Finally, let us simplify the expressions somewhat by assuming that the situation is 
truly isotropic. Then one would expect\footnote{The results of the evolutions carried out
by \citet{schw88} suggest that the isotropic assumption is accurate to 10-20\% for $I_\parallel$. 
At the same time $I_l \neq0$. Give the many other uncertainties associated with the neutron star problem, it seems 
reasonable to assume that the turbulent tangle is isotropic here.}
  $I_\parallel = 2/3$ and $I_l=0$. This leaves us with 
the force expression
\be
{f}_i^\mathrm{mf} = { 2 \over 3} \kappa \rho_\n { { \cal B} c_\L^2 \over \tilde{\nu}^2} w_{\n\p}^2 {w}_i^{\p\n} 
\ee

If we compare this expression to (\ref{vinen}) we see that the two expressions would be identical 
if 
\be
c_\L = { 2 \pi \over \kappa} \left( {\chi_1 \over \chi_2} \right) {\cal B} \tilde{\nu}
\ee

This analysis highlights the central r\^ole of the averaging, i.e. the integration over all vortex segments in the 
sample volume. In order to be able to analyse less symmetric situations one would have to devise a practical way of carrying
out the integration.

\section{Digression: Potential implications for neutron stars}

It is interesting to pause at this point and discuss the implications 
that a mutual friction force of form (\ref{vinen}) might have for neutron star dynamics.
It is natural to first compare the relative strength of the alternative 
mutual friction forces.  In the case of a straight vortex array we have 
the Hall-Vinen type force
\be
f_\mathrm{HV} \approx \mathcal{B} \rho_\n \kappa n_v w_{\p\n} 
\ee
per unit volume.
Meanwhile, in the case of isotropic turbulence we have argued that we should use
the Gorter-Mellink force
\be
f_\mathrm{GM} \approx { 8 \pi^2 \over 3 \kappa} \rho_\n \left( {\chi_1 \over \chi_2} \right)^2 \mathcal{B}^3 w^3_{\p\n}
\ee
The relative magnitude of these forces is  
\be
{ f_\mathrm{GM} \over f_\mathrm{HV}} 
\approx { 8 \pi^2 \over 3} \left( {\chi_1 \over \chi_2} \right)^2 { \mathcal{B}^2 w^2_{\p\n} \over n_v \kappa^2} 
 \ee
Taking $\chi_1/\chi_2 \sim 1$ and recalling that $n_v\kappa \approx 2 \Omega_\n$, where $\Omega_\n$ is 
the rotation rate of the superfluid, we readily arrive at 
\be
{ f_\mathrm{GM} \over f_\mathrm{HV}} 
\approx {4 \pi^2 \over 3}  \mathcal{B}^2 { w^2_{\p\n} \over \kappa \Omega_\n} \approx 
10^9 \left( { r \over 10^6 \mathrm{ cm} } \right)^2 \left( { \Delta \Omega \over \Omega_\n} \right)^2 \left( { \Omega_\n \over 1 \mathrm{ rad/s}
} \right) 
\ee
Here we have used $w_{\p\n} = r \Delta \Omega = r (\Omega_\p - \Omega_\n)$ (with $r$ the radial distance from the rotation axis)
and $\kappa = h/2m_n \approx 2\times 10^{-3}$~cm$^2$/s, together with the typical value 
$\mathcal{B}\approx 4\times10^{-4}$.
Finally, as a representative value of the relative rotation rate attainable in a neutron star we take 
$\Delta \Omega/\Omega_\n \approx 5 \times 10^{-4}$ which is suggested by pulsar glitch data \citep{lyne}.
This leads to the final estimate
\be
{ f_\mathrm{GM} \over f_\mathrm{HV}} \approx 250 \left( { r \over 10^6 \mathrm{ cm} } \right)^2 
\left({\Delta \Omega/\Omega_\n \over 5 \times 10^{-4} }  \right)^2
\left( { P \over 1 \mathrm{ s} } \right)^{-1} 
\ee
where $P$ is the rotation period. This estimate shows that the relative strength of the two forces depends on the location in the star.
Near the rotation axis the turbulent mutual friction would be significantly weaker than the  force  for
a straight vortex array. Meanwhile, far way from the axis, eg. near the equator in the outer neutron star core, the turbulent force may actually
be stronger than the standard result. That the relative magnitude of the forces is location dependent is obvious since 
they depend on the relative velocity in different ways.   
It should be noted that our estimate for the relative strength of the two forces differs significantly from 
that of \citet{peralta1}. They argue that the turbulent force is five orders of magnitude weaker than the 
straight vortex force. The difference is simply due to Peralta et al using a smaller value of the relative velocity in their estimate.
 
Next let us discuss the implications of fully developed superfluid turbulence for the dynamical 
coupling of the two fluids in a neutron star core. We have previously shown \citep{trevor} that the standard mutual friction coupling leads to 
damping of any relative rotation on a timescale of 
\be
\tau_d \approx 10 P (\mathrm{s})
\left( { m_\p^* \over \delta m_\p^*} \right)^2  \left( {x_\p \over 0.05} \right)^{-1/6} \left({ \rho \over 10^{14} \mathrm{g/cm}^3} 
\right)^{-1/6} \ .
\label{relax}\ee
In our discussion, we noted that 
this estimate is about
one order of magnitude smaller than the classic result of \citet{as88}. We argued that the difference 
was due to the different scenarios considered. Further consideration of the problem shows that, if one imposes 
global conservation of angular momentum in the \citet{as88} scenario then one arrives at a coupling timescale very similar to ours. 
This point has, in fact, already been made by \citet{jahanmiri}. Hence, we conclude that our estimate of the coupling timescale is 
reliable.

Let us now consider the turbulent mutual friction force. As in \citet{trevor} one can show that 
the relative velocity evolves according to 
\be
\partial_t w_i^{\n\p} = { 1 \over x_\p} \left( { m_\p \over m_\p^*} \right) { f_i^\mathrm{mf} \over \rho_\n}
\ee
In the present case this leads to (with $w_{\n\p}\to w$ for clarity) 
\be
\partial_t w \approx A w^3 \rightarrow  w = { w_0 \over \sqrt{ 1 + A w_0^2 t} } 
\ee
where
\be
A = { 8 \pi^2 \over 3 \kappa} { 1 \over x_\p} \left( { m_\p \over m_\p^* } \right) \left( { \chi_1 \over \chi_2} \right)^2 {\cal B}^3  
\ee
and $w_0=w(t=0)$. This behaviour is clearly rather different from that infered in the standard case. In particular, 
it is clear that the coupling is typically much slower than exponential. In fact, it is easy to show that there may be a regime
where the relaxation is essentially linear (at least locally). This would be the case if 
\be
t \ll t_\mathrm{linear} \ , \qquad \mbox{ where } \qquad t_\mathrm{linear} = { 1 \over A w_0^2 } 
\ee
For typical parameters (the same as in (\ref{relax})) we find 
\be
t_\mathrm{linear} \approx 5 \times 10^{-3} \left( { \chi_1 \over \chi_2} \right)^{-2} \left( {x_\p \over 0.05} \right)
\left( {P \over 1 \mathrm{ s} }\right)^2 \left( { r \over 10^6 \mathrm{ cm}} \right)^{-2} \mathrm{ s}
\ee 
This suggests that a region near the rotation axis of the star (within a metre or so) will relax linearly
on a timescale of days. This behaviour is certainly sufficiently different from the standard exponential
relaxation to warrant further investigation.

\section{Polarised turbulence}

At this point we have achieved what we originally set out to do. We have generalised the mutual friction force to 
account for the fact that the vortices may not be straight. We have also shown how one can ``derive'' the force that would be appropriate 
when   the vortices form an isotropic tangle and the superfluid exhibits fully developed turbulence.
Our results are essentially taken over from a number of studies of the analogous problem for Helium. 
However, as we  developed the model it  became clear that the two cases we have so far considered
represent extremes rather than the generic situation. In particular, the approach to superfluid turbulence 
that we adopted is based on ``counterflow'' experiments in Helium \citep{donnelly}. In this context, all
vortices are excitations generated by the flow and turbulence sets in above a critical relative velocity. 
This is in sharp contrast to the neutron star problem, where a dense vortex array representing the bulk rotation 
would be present from the outset. In this case the relevance of the turbulent mutual friction force (\ref{vinen}) is
far from clear. What is clear is that, if one assumes fully developed turbulence then there can be no large
scale 
rotation in the superfluid. After all, we assumed that the vortex tangle is isotropic. On the macroscopic 
scale, each ``fluid element'' would then have to be irrotational. Hence, results obtained by simply replacing
(\ref{fmf1}) by (\ref{vinen}) are debatable. This casts some doubt on the relevance of the recent simulations carried out by 
\citet{peralta1,peralta2}. On the other hand, it is entirely possible that turbulence will develop locally in a neutron star, 
perhaps in connection with a pulsar glitch. One should certainly consider the implications of this further. 
Yet it is also clear that we need to add another dimension to our analysis. 

\subsection{The Helium problem}

In recent work on superfluid Helium, the possible interaction between relative flow and rotation
has been considered. It is probably fair to say that this problem is still not well understood. 
There certainly does not yet exist an agreed upon model that we can adapt to
suit the neutron star problem. Nevertheless, there are useful hints. Much of the recent work was inspired by 
an experiment carried out by \citet{swanson}. The experiment concerned counterflow in a rotating cylindrical vessel (heated from below) 
and the results can be summarised as follows. 
In absence of rotation,  turbulence sets in above a critical counterflow rate. Representing the relative flow by $W$, 
one finds that above a certain critical value $W_c$ the vortex line density grows as $L \sim W^2$.
Once rotation is considered, it appears that the original
$W_c$ is eliminated already at very slow rotation rates. Instead, \citet{swanson} find that i) for low values of $W$ 
there is a flat region in $L$, ii) at a first critical flow $W_{c1}$ there is a slight rise to a second flat region, and iii)
beyond a second critical flow rate $W_{c2}$ one sees the anticipated rise with $L\sim W^2$. Nevertheless, 
rotation appears to add fewer than the anticipated $L=L_R=2\Omega/\kappa$ lines.

Simulations (analogous to those of \citet{schw88}) aimed at understanding the experimental results have been carried out by
\citet{tsubota1} and \citet{tsubota2}. They confirm much of the phenomenology seen in the experiment. 
The analysis associates the lowest critical velocity $W_{c1}$ with the onset of the so-called Donnelly-Glaberson instability \citep{glaberson}. 
This means that one would have $W_{c1}\approx W_{DG} = 2 (2\nu \Omega)^{1/2}$.
\citet{tsubota2} also observes that, even at high relative flow rates, the disordered tangle 
has roughly the same ``rotation'' as the initially ordered array. This makes some sense since the overall angular momentum 
should be conserved. Of course, it also suggests that the mutual friction force will not simply be described by (\ref{vinen})
in a rotating system. Instead, one should consider a \underline{polarised} turbulent state, where the force is essentially
 a linear combination of (\ref{mf_macro}) and (\ref{vinen}) 

To get some idea of how this may work out, we consider the phenomenological model suggested by \citet{jou1}.
They generalise Vinen's equation by noting that, when both rotation and relative flow are considered then 
the problem has two dimensionless parameters, $W/\kappa L^{1/2}$ and $(\Omega/\kappa L)^{1/2}$. Including terms up to 
quadratic order, but neglecting $W^2$ in anticipation of the relative flow being modest, 
they write down the following evolution equation for the line density, 
\be
{dL \over dt} = - \beta \kappa L^2 + ( \alpha_1 W + \alpha_2 \sqrt{\kappa \Omega}) L^{3/2} - \left( \beta_1 \Omega + \beta_2 W \sqrt{\Omega\over \kappa} \right)L
\label{jmeq}
\ee
The coefficients $\alpha_1$ and $\beta$ follow from the original Vinen description, 
but the other coefficients are new and describe the competition between relative
flow, which tends to drive the turbulence, and overall rotation, that has a stabilising influence. To find the stationary 
state of the system we set $dL/dt=0$. This  leads to a simple quadratic for $L^{1/2}$. This quadratic is obviously straightforward to solve
in general. 
Nevertheless, in order to reduce the number of free parameters in the model, and simplify the solutions, it is useful to impose the constraint 
\be
{ \beta_1 \over \beta} = { \beta_2 \over \alpha_1} \left( { \alpha_2 \over \beta} - {\beta_2 \over \alpha_1} \right)
\label{const}\ee
Then the two roots become
\be
L^{1/2} = { \beta_2 \over \alpha_1} \sqrt{ \Omega \over \kappa}
\ee 
and
\be
L^{1/2} = { \alpha_1 \over \beta} { W \over \kappa}  + \left( { \alpha_2 \over \beta} - { \beta_2 \over \alpha_1} \right) \sqrt{\Omega\over \kappa}
= { \alpha_1 \over \beta} \left[{ W \over \kappa} + {\beta_1 \over \beta_2} \sqrt{\Omega \over \kappa} \right]
\label{sol2}\ee
Analysing the stability of these solutions, one finds that there exists a critical relative velocity
\be
W_c = \left( {\alpha_2 \over \alpha_1} - 2 { \beta_1 \over \beta_2} \right) \sqrt{ \kappa \Omega} 
\ee
The first solution is stable below this velocity, while the second solution is stable above it.

\begin{figure}
\begin{center}
\includegraphics[width=8cm,clip]{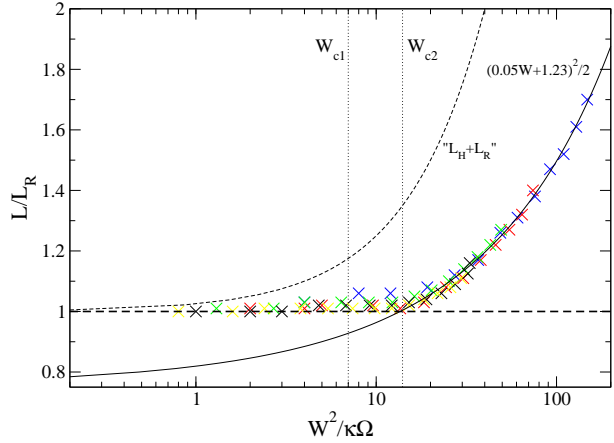} 
\end{center}
\caption{The experimental results of \citet{swanson}, read off from their Figure~3 and represented by crosses in the figure, are compared to the phenomenological model of
\citet{jou1} discussed in the text. The determination of the various parameters in the model is discussed in Appendix~B. 
Clearly, the model (given by the dashed line below $W_{c2}$ and the solid curve above it) describes the experimental results quite well.}
\label{helium}\end{figure}

This relatively simple model turns out to describe the experimental results quite well. There is 
a flat region in $L$ below $W_c$ and 
the anticipated quadratic behaviour above $W_c$. Reading off the experimental data from Figure~3 in \citet{swanson} (see also Appendix~B) 
and rescaling to our dimensionless quantities
we find that the results for different rotation rates are well represented by the model, see Figure~\ref{helium}.
This is, in fact, a useful sanity check since
it shows that the various coefficients do not depend on $W$ or $\Omega$ in a significant way.
The figure also shows that, taking $L_H^{1/2}={ \alpha_1 \over \beta} { W \over \kappa}$, we have 
$L<L_H+L_R$. 
The model cannot account for the small step 
in $L$ observed at $W_{c1}$ but for our present purposes this may not be very important.

Given the apparent success of this relatively simple model it would be interesting to apply it to the neutron star problem.
Then we first of all need to be able to determine the various new coefficients. To arrive at some reasonable values
is relatively easy. First compare (\ref{jmeq}) to (\ref{vinen}) to see
that 
\bea
\alpha_1 &=& \chi_1 {\cal B} \\
\beta &=& { \chi_2 \over 2 \pi} 
\eea 
Next consider the slow-rotation limit where one would expect the vortex array to be straight. Then $L=L_R$ which means that
\be
\beta_2 = \sqrt{2} \alpha_1 
\ee
The final two coefficients are obtained from (\ref{const}) and the identification of $W_c$ with the critical velocity for the onset of the
Donnelly-Glaberson instability\footnote{We are currently considering this instability in the neutron star case. Our preliminary results suggest that the
critical velocity remains the same, which would make sense.} $W_{DG}= 2 (2\nu\Omega)^{1/2}$. This leads to 
\bea
\beta_1 &=& { \chi_2 \over \pi} - 4 \left({ \nu \over \kappa} \right)^{1/2} \chi_1 {\cal B} \approx 2 \beta
\\
\alpha_2 &=& \sqrt{2} \left[{ \chi_2 \over \pi} - 2 \left({ \nu \over \kappa} \right)^{1/2} \chi_1 {\cal B} \right]
\approx 2 \sqrt{2} \beta
\eea
since ${\cal B}\ll1$.  

\subsection{A phenomenological approach to the neutron star problem}

Now that we have a basic picture of the combined problem of rotation and relative flow, we 
can try to formulate a prescription for turbulence in a rotating superfluid neutron star. By necessity, this will 
require a considerable leap of faith. After all, the Helium problem is not completely understood and 
in that case one has the luxury of being able to carry out laboratory studies. What hope do we have of 
making progress on the neutron star problem? We believe there are reasons to be optimistic. 
After all, a new perspective on superfluid dynamics is emerging and it will be quite exciting 
to see if even a very basic phenomenological model can help us understand observed phenomena, eg. related
to glitches.   

We want to work at the macroscopic fluid dynamics level, combining our 
description of mutual friction for a straight vortex array (\ref{fmf1}) with
that for a vortex tangle (\ref{vinen}) in a ``sensible'' way. 
To do this we will assume that  we can average over the collection of vortices
in each fluid element. In the presence of a turbulent vortex tangle we will assume that
the averaging can be done on a lengthscale large enough that each fluid element contains an isotropic 
tangle as well as a ``polarised'' set of vortices that can be represented by a (locally) straight
vortex array\footnote{Note: This approach may fail unless there is a distinguishable upper limit on the size of the largest vortex loops.
Having said that, it is clear from the beginning that the separation into straight and tangled vortices in each fluid element is artificial.
The assumption is made in order to complete the model. To what extent it is sensible depends on future results concerning, for example, post-glitch dynamics. 
At present we are simply trying to develop a model for which ``we can do calculations''.  }. 

Combining what we have learned from the Helium problem, a ''workable'' strategy 
might be as follows. First note that the turbulent vortex tangle does not 
contribute to $\nabla \times \vec{v}_\n$, while the polarised piece obviously does. 
Neglecting entrainment, which is difficult to account for in this phenomenological model, we let the average 
\be
\langle \nabla \times \vec{v}_\n \rangle = 2 \vec{\omega}_\n = n_v \vec{\kappa}
\ee
define the vorticity $\vec{\omega}_\n$ on the macroscopic scale.
Taking the absolute value of this we can then identify
\be
| \langle \nabla \times \vec{v}_\n \rangle | = 2 \omega_\n = L_R \kappa
\ , \quad \mbox{ that is } \quad L_R = { 1 \over \kappa} | \langle \nabla \times \vec{v}_\n \rangle |
\ , \quad \mbox{ and} \quad n_v = L_R
\ee
We simply associate the vortex line length due to the local rotation $L_R$ with the 
density of straight vortices per unit area.  Finally we make 
contact with the model discussed in the previous section by taking $\Omega=\omega_\n$.

Next we need the relative flow $W$ in (\ref{jmeq}). In the Helium experiment this is
the counterflow along the rotation axis. Hence it seems natural to represent it by the projection of the relative flow
$w^{\n\p}_i$ along the local rotation ``axis''. This means that we would have
\be
W = | \vec{w}_{\n\p} \cdot \hat{\omega}_\n | =  | \vec{w}_{\n\p} \cdot \hat{\kappa} | 
\ee
Here we have assumed that $W>0$. This is not a physical assumption, in fact, there is no reason 
why the flow should not be in a direction opposite to the rotation axis. However, we need to introduce this restriction to 
keep contact with (\ref{jmeq}).

Now we have all the ingredients required to work out the total line length $L$ per unit volume. 
First we should check if the relative flow exceeds the critical velocity for the onset of turbulence
$W_c$. If it does not, then the sample is not turbulent and it is appropriate to use the Hall-Vinen
form for the mutual friction (\ref{fmf1}). If, on the other hand, $W>W_c$ then we need a prescription for
working out $L$. Provided that we can assume that the system is in ``equilibrium'' this is easy, and 
the required solution is given by  (\ref{sol2}). From this solution we readily deduce the total line
length in the turbulent tangle $L_T = L-L_R$. Explicitly, we would have 
\be
L_T = \left( {\alpha_1 \over \beta} { W \over \kappa}\right)^2
+ { \alpha_1 \beta_1 \over \beta^2} {W \over \kappa} L_R^{1/2} + \left[ { 1 \over 4} \left( {\beta_1 \over \beta}\right)^2 -1 \right]L_R
\label{lteq}\ee
This allows us to work out the mutual friction due to the tangle by using $L_T$ in (\ref{fturb1}).
Finally, the contribution from the polarised part follows  from combining (\ref{fmf1}) 
with the deduced vortex density $n_v =L_R$.
The combined mutual friction force for the polarised turbulent state (obviously only relevant above the critical velocity) can then be written 
\be
f_i^{\rm mf} = \rho_\n L_R \left\{ {\cal B}^\prime  \epsilon_{i j k} \kappa^j 
                   w_{\n\p}^k + {\cal B} \epsilon_{i j k} 
                   \epsilon^{k l m} \hat{\kappa}^j \kappa_l w^{\n\p}_m
-  \tilde{\nu} \left[ {\cal B}^\prime \hat{\kappa}^j \nabla_j \kappa_i 
+    {\cal B} \epsilon_{ijk} \kappa^j \hat{\kappa}^l \nabla_l \hat{\kappa}^k \right] \right\} +  { 2 L_T \over 3} \rho_\n \kappa {\cal B} w^{\p\n}_i  
                   \ .
\ee

This model represents what we think may be the simplest ``reasonable'' approach to the problem of 
superfluid neutron star turbulence. It has all the key ingredients, combining 
the standard straight vortex mutual friction with a contribution from an isotropic turbulent tangle.
The main additional advantage is that all the steps can actually be worked out in a practical situation. 
Of course, the model is phenomenological and comes with a number of (potentially serious) caveats. 
Obviously, when we write down (\ref{lteq}) we assume that the system is in 
equilibrium. This does not have to be the case. If one expects the timescale for reaching 
equilibrium to be long, then one can instead assume that $L$ evolves according to (\ref{jmeq}).    
The assumption that $W>0$ may also need to be relaxed. This can be done by considering the directionality in the problem in more detail.
There have been some efforts to do this in the Helium case, see for instance \citet{nem}, \citet{lip} and \citet{jou2}.
Common to the ideas discussed in these papers is the use of the Onsager symmetry principle to ensure that the
second law of thermodynamics is honoured. This is important since the system is dissipative. An interesting approach 
to the problem was outlined in a series of papers by  \citet{geu88,geu89,geu92} and \citet{geu94, geu95, geu96}. 
The formalism developed in these papers seems particularly promising given that the equations
for the basic two-fluid system are derived from a variational principle similar to that used by \citet{prix}.
An additional ``fluid'', represented by the scalar density $L$
and the flow $\vec{v}_L$,  is introduced to describe the turbulent tangle. This seems like a natural approach to take and there could be a considerable overlap
with the general framework for dissipative multi-fluid systems recently developed by two of us \citep{monster}. 
We are planning to investigate this possibility in detail in the future. 

\section{Concluding remarks}

We have discussed the problem of turbulence and the associated mutual friction force in a superfluid neutron star core. 
Our analysis was heavily influenced by work done on the analogous problem for superfluid Helium over the last 50 years, starting with the seminal work of \citet{feynman}
and \citet{vinen57}. 
Focussing on the aspects of the problem that should be relevant for neutron stars, 
we have derived the form of the mutual friction force that should apply 
if the system becomes completely turbulent. We then argued that the neutron star case is likely to require an understanding 
of polarised turbulence, where rotation has a stabilising influence on the ability of relative flow to induce turbulence.
In recognition of this, we formulated a phenomenological model that represents a first attempt (for neutron stars) to 
combine the presence of local rotation (represented by a ``straight'' vortex array'') and turbulence (an isotropic 
vortex tangle). 

These results are intriguing, and possibly quite important. It will certainly be interesting 
to consider the effects that turbulence may have on, for example, pulsar glitch relaxation and neutron star oscillations.  
Having said that, we appreciate that this was our very first attempt at understanding the turbulence 
problem. The analysis may have taken us some way towards 
the solution to the problem, but more than likely we are only beginning to understand 
what the real questions are.

\section*{Appendix A: The induced flow for a curved vortex} 

\begin{figure}
\begin{center}
\includegraphics[width=8cm,clip]{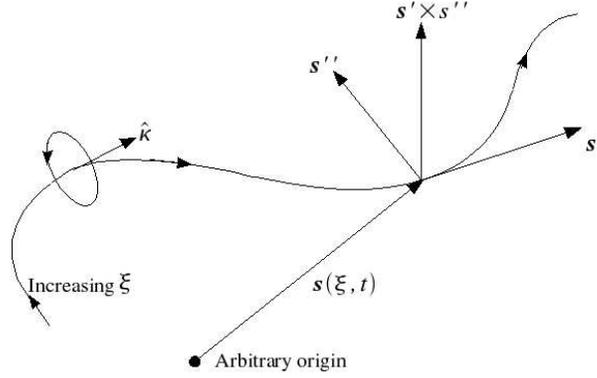} 
\end{center}
\caption{A schematic illustration of the variables used to describe the self-induced 
flow of a superfluid vortex. }
\label{curved}\end{figure}

Let us focus on a single vortex. In other words, we assume
that the vortices are well separated, which implicitly puts some restrictions 
on the vortex curvature. 
To describe the vortex we take $s_{i}(\xi,t)$, 
where $\xi$ is a parameter denoting position along the vortex 
(essentially the arc length), as the position vector of 
the vortex from an arbitrary origin.
The situation is illustrated in Figure~\ref{curved} and it is useful to 
note that
\bea
s_i^\prime &=& {ds_i \over d\xi} = \hat{\kappa}_i \\
s_i^{\prime \prime} &=&  {d^2s_i \over d\xi^2} = \s^{\prime j} \nabla_j \s_i^\prime
\eea 
A prime represents a derivative with respect to $\xi$. 
It is also useful to note the alternative form
\be
s_i^{\prime \prime} = - \epsilon_{ijk} s^{\prime j} \epsilon^{klm} \nabla_l s^\prime_m
\ee

To determine the self-induced velocity  we must solve for the 
flow associated with the vortex. To do this we first note that it is the 
circulation of momentum $\bar{p}_i$ that is quantised (as in the main text, we distinguish the mesoscopic 
momentum by a bar), not that of velocity. As discussed in detail 
by \citet{trevor} this is an important distinction to make once we include the entrainment. 
We need to solve (omitting for the moment any constituent indices)
\begin{equation}
{1 \over m} \epsilon_{ijk}\nabla^{j} \bar{p}^k = \kappa_i = { h \over 2 m} \hat{\kappa}_i
\end{equation}
We can write down a formal solution in terms of a Green's function 
by  introducing a vector potential such that
\begin{equation}\label{eq:DefnA} 
\epsilon_{ijk}\nabla^{j} {A}^{k} = { 1 \over m} \bar{p}_{i}
\end{equation}
Using the associated gauge freedom to set
\begin{equation}
\nabla^{j} {A}_{j} = 0
\end{equation}
we see that we need to solve a vector Poisson equation;
\begin{equation}
 \nabla_{j} \nabla^{j} {A}_{i} = - {\kappa}_{i}
\end{equation}
The formal solution is given by
\be
A_i (\vec{x}) = { 1 \over 4\pi} \int { \kappa_i \over | \vec{x} - \vec{y}| } d^3 y
\ee
From this it follows that
\begin{equation}
{\bar{p}_i(\vec{x}) \over m} = - { 1 \over 4\pi} \int {\epsilon_{ijk} R^j \kappa^k \over R^3} d^3 y 
\end{equation}
where $R^i = x^i - y^i$. To make use of this solution we note that $\kappa^i$ has support only on the vortex. 
Thus the volume integral collapses to a line integral along the vortex, in terms of $\xi$. Then we have
$R^i = x^i - s^i$, and assuming that the vortex is only slightly bent we can make use of the Taylor expansion
\be
s_i \approx s_i^0 + s_i^\prime \xi + { 1 \over 2} s_i^{\prime \prime} \xi^2 
\ee
This also leads to 
\be
\hat{\kappa}_i \approx s_i^\prime + s_i^{\prime \prime} \xi
\ee
Finally, we note that if we want to work out the corrections to the Magnus effect 
decribed by \citet{trevor} we are interested in the induced flow in the vicinity of the vortex. 
This means that we take the limit $x^i \to s_0^i$, to arrive at
\be
\epsilon_{ijk} R^j \hat{\kappa}^k \approx 
- { 1 \over 2} \epsilon_{ijk}s^{\prime j} s^{\prime\prime k} \xi^2 
\ee
and
\be
{\bar{p}^\mathrm{ind}_i \over m} = { \kappa \over 8 \pi} \int { \epsilon_{ijk}s^{\prime j} s^{\prime\prime k}  \over \xi} d \xi
\ee
To carry out the integration we introduce a cut-off at $\xi = a_0$, taken to be the size of the vortex core. This regularises the 
integral and also makes sense since we have not made any assumptions about the nature of the vortex core. We also 
introduce an upper limit $\xi=b$, 
assuming that the contribution to the induced flow can by localised to some part of the vortex. As a resonable 
value for $b$ one can use the intervortex separation. Finally, we arrive at the 
expression
\be
{\bar{p}^\mathrm{ind}_i \over m} = { \kappa \over 4 \pi} \log \left( { b \over a_0} \right) \epsilon_{ijk}s^{\prime j} s^{\prime\prime k} \equiv
\nu \epsilon_{ijk}s^{\prime j} s^{\prime\prime k} 
\ee
In deriving this result we assumed symmetry with respect to $\xi=0$.

\section*{Appendix B. Determining the parameters in the phenomenological Helium model}

The parameters  used in Figure~\ref{helium} were arrived at in the way discussed at the end of section~7.1.
Specifically, we first take $L=L_R$ below the critical velocity. This fixes
\be
 { \beta_2 \over \alpha_1} = \sqrt{2}
\ee 
A rough fit  to the experimental results allows us to determine, cf. (\ref{sol2}), 
\be
{ \alpha_1 \over \beta} \approx 5 \times 10^{-2} 
\ee
and
\be
{ \alpha_ 1 \over \beta } { \beta_1 \over \beta_2} \approx 1.23 \rightarrow { \beta_1 \over \beta_2} \approx 24.6
\ee
We can now combine these different estimates to get 
\be
{ \beta_2 \over \beta} \approx 0.071
\ee
and then
\be
{ \beta_1 \over \beta} \approx 1.75
\ee
Next we assume that the critical velocity $W_{c2}$ corresponds to $W_{DG}$. This means that 
we should have $W_{c2}= 2 \sqrt{2 \nu \Omega}$, or
\be
{ \alpha_2 \over \alpha_1} - 2{ \beta_1 \over \beta_2} = 2 \sqrt{ 2\nu \over \kappa}
\ee
Combining this with  the observed critical velocity $W_{c2} \approx 0.118 \sqrt{\Omega} $~cm/s$^{1/2}$ we find that
\be
{ \alpha_2 \over \beta} \approx 2.65 
\ee
We now have all parameters in (\ref{jmeq}) expressed in terms of $\beta$. The values we have 
deduced agree quite well with those used by \citet{jou1}. Finally, we learn from \citet{vinen57} that 
$\chi_2 \approx 1$ which means that we could use
\be
\beta = { \chi_2 \over 2\pi} \approx 0.16
\ee
Then all parameters in the model have been specified. 

\section*{Acknowledgements}

We would like to thank Carlo Barenghi, Andrew Melatos and Carlos Peralta for interesting discussions
of this problem. 
NA acknowledges support from PPARC via grant no PPA/G/S/2002/00038 and Senior Research Fellowship no
PP/C505791/1. GLC is supported by NSF grant no PHY-0457072.

\end{document}